\documentclass[pre,twocolumn,aps,showpacs,floatfix]{revtex4}
\usepackage{graphicx}
\usepackage{dcolumn}
\usepackage{bm}

\begin{document}

\title{Artificial photosynthetic reaction centers coupled to light-harvesting antennas}

\author{ Pulak Kumar Ghosh$^{1}$,  Anatoly  Yu. Smirnov$^{1,2}$, and Franco Nori$^{1,2}$}

\affiliation{ $^1$ Advanced Science  Institute, RIKEN, Wako-shi, Saitama, 351-0198, Japan \\
$^2$  Physics Department, The University of Michigan, Ann Arbor,
MI 48109-1040, USA }

\date{\today}

\begin{abstract}
We analyze a theoretical model for  energy and electron transfer in
an artificial photosynthetic system. The photosystem consists of a
molecular triad (i.e., with a donor, a photosensitive unit, and an
acceptor) coupled to four accessory light-harvesting antennas
pigments. The excitation energy transfer from the antennas to the
artificial reaction center (the molecular triad) is here described
by the F\"{o}rster mechanism. We consider two different kinds of
arrangements of the accessory light-harvesting pigments around the
reaction center. The first arrangement allows direct excitation
transfer to the reaction center from all the surrounding pigments.
The second configuration transmits energy via a cascade mechanism
along a chain of light-harvesting chromophores, where only one
chromophore is connected to the reaction center. At first sight, it
would appear that the star-shaped configuration, with all the
antennas directly coupled to the photosensitive center, would be
more efficient. However, we show that the artificial photosynthetic
system using the cascade energy transfer absorbs photons in a
broader wavelength range and converts their energy into electricity
with a higher efficiency than the system based on direct couplings
between all the antenna chromophores and the reaction center.
\end{abstract}

%\pacs{87.16.A-, 87.16 Tb, 73.63.-b}

\maketitle

\section{Introduction}

The reaction centers of natural photosystems are surrounded by a
number of accessory light-harvesting complexes \cite{Alberts}.
These light-harvesting antennas absorb sunlight photons and
deliver their excitation energy to the reaction center, which
creates a charge-separated state. The photosystem of green plants
is made up of six photosynthetic accessory pigments: carotene,
xanthophyll, phaeophytin $a$, phaeophytin $b$, chlorophyll $a$,
and chlorophyll $b$ \cite{Alberts}. Each pigment absorbs light in
a different range of the solar spectrum. As a result, the antenna
complex significantly increases the effective frequency range
 for the light absorption,  resulting in a highly-efficient
 photocurrent generation. In the presence of
excessive sunlight the antenna complex can reversibly switch to
the photo-protected mode, where harmful light energy is
dissipated.

%\end{document}
%%%%%%%%%%%%%%%%%%%%%%%%%%%  Figure 1  %%%%%%%%%%%%%%%
\begin{figure}[tp]
\centering
\includegraphics[width=8.6cm]{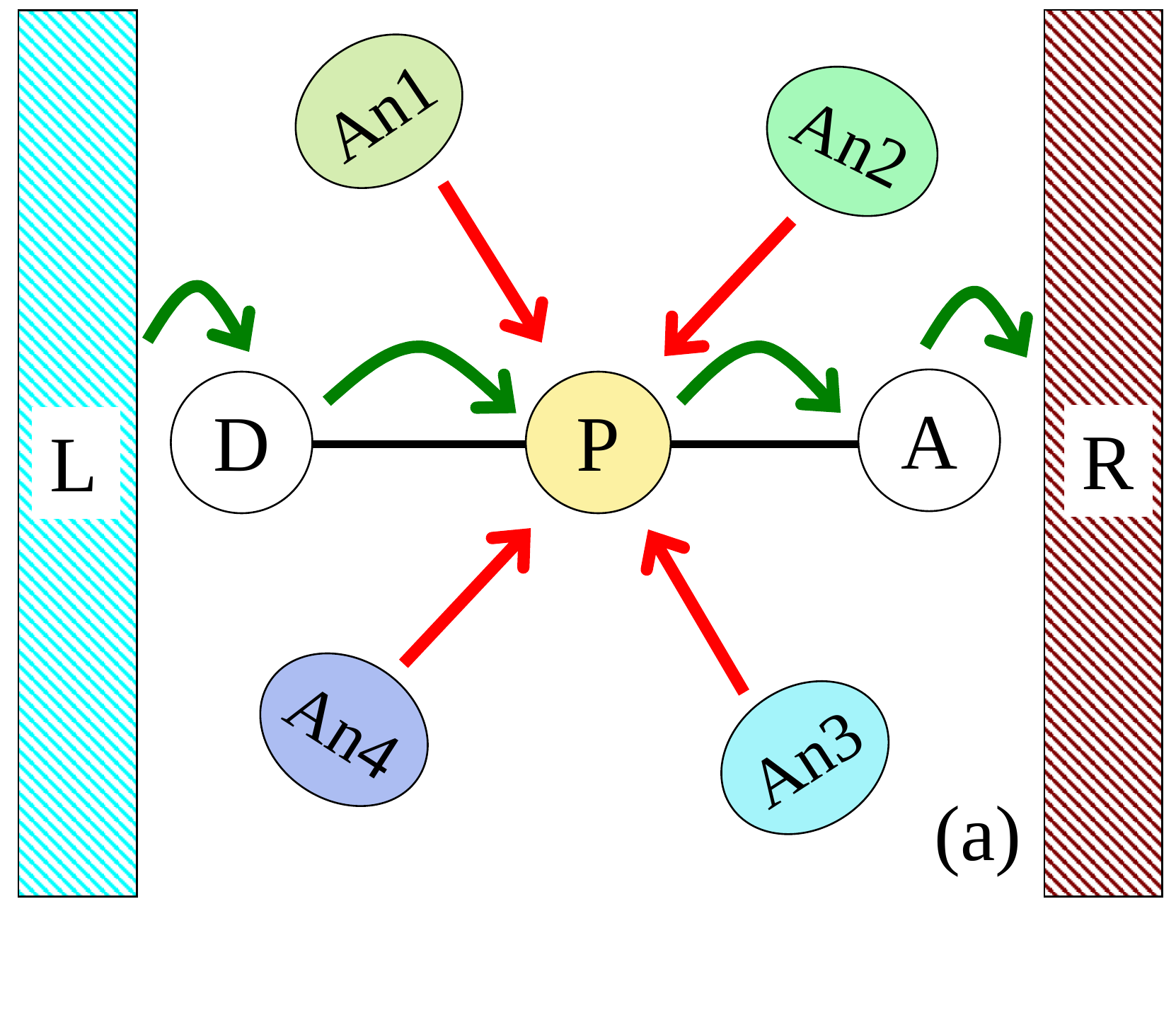}
\includegraphics[width=8.7cm]{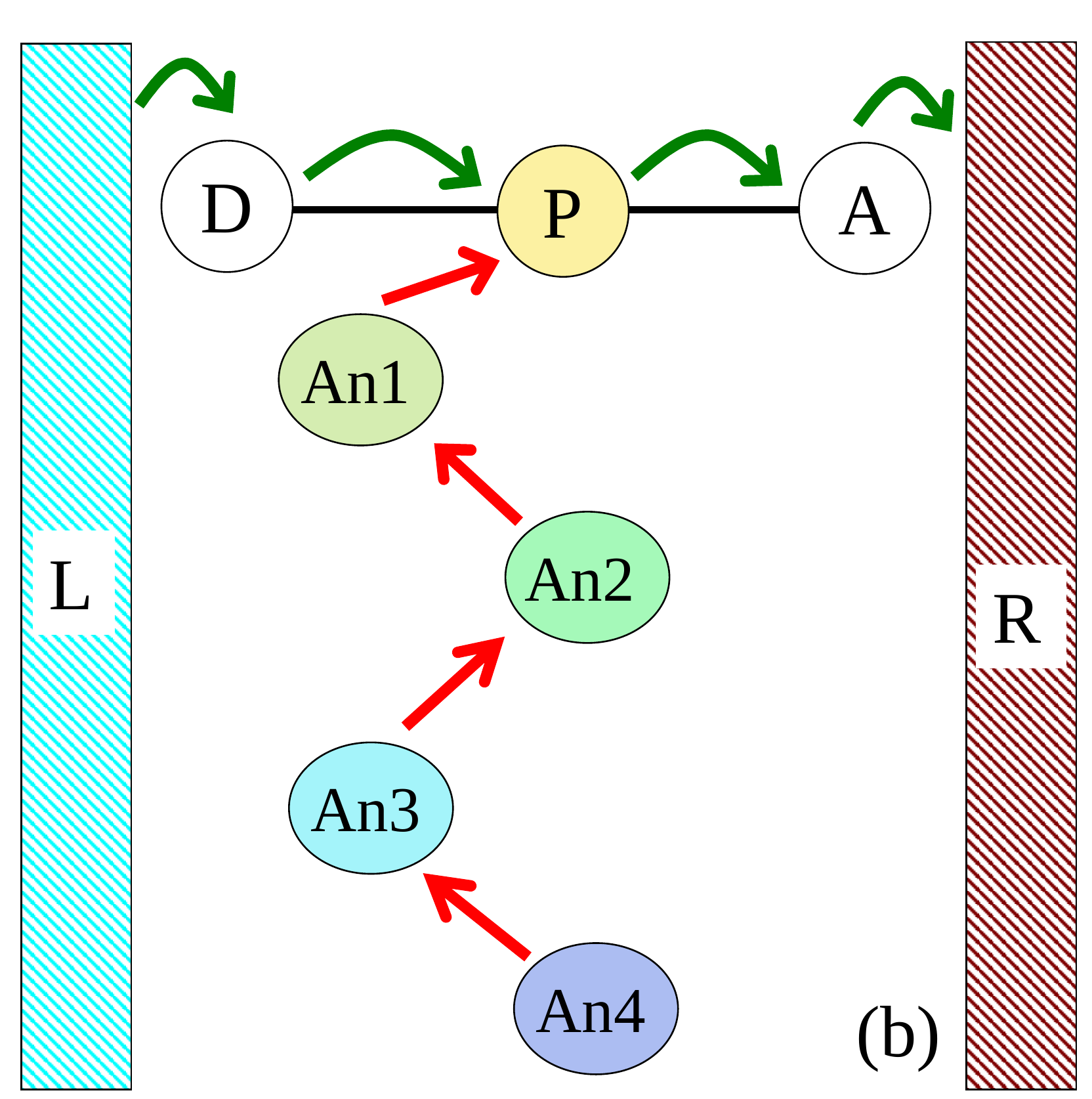}
\caption{(Color online) Schematic diagram of an artificial
photosystem comprised of a molecular triad (D--P--A) and four
additional light-harvesting complexes (An1, An2, An3, An4). Here,
D = donor, A = acceptor, and P = photosensitive part (porphyrin).
The molecular triad D--P--A is inserted between two electrodes
(leads) L and R. Energy exchange processes are denoted by straight
red arrows. The green bent curved arrows describe electron
pathways, L $\rightarrow$ D $\rightarrow$ P $\rightarrow$ A
$\rightarrow$ R, via the molecular triad. (a) The photosensitive
part, P, of the molecular triad is surrounded by four accessory
light-harvesting complexes An1, An2, An3, and An4. In this case
the surrounding antenna complexes can transfer excitations to the
reaction center directly. (b) Here the antenna complexes form a
linear chain coupled to the reaction center via nearest-neighbor
couplings. \label{F1}}
\end{figure}
%%%%%%%%%%%%%%%%%%%%%%%%%%%%%%%%%%%%%%%%%%%%%%%%%%%%%%%%%%%%%%%

 The efficient performance of natural photosystems motivates
researchers to mimic their functions by creating photosynthetic
units, which are combined to antenna complexes with artificial
reaction centers. For example, a light-harvesting array of
metalated porphyrins has been developed in Ref.~\cite{gust1}. This
array absorbs light and rapidly transfers the excitation energy to
the reaction center, so that the porphyrin (P)--fullerene
(C$_{60}$) charge-separated state, P$^{+}$--C$_{60}^{-}$, is
formed with a quantum yield $\sim 70\%$. Mixed self-assembled
monolayers of the ferrocene (Fc)--porphyrin--fullerene molecular
triad and the boron dipyrrin (B) dye have been made in
Ref.~\cite{imahori1, imahoriR} with the goal to examine both
energy and electron transfers in the artificial reaction center
(Fc--P--C$_{60}$), coupled to the light-harvesting molecule B. A
quantum yield of $\sim 50\%$ for photocurrent production at the
wavelength 510 nm and a quantum yield of $\sim 21 \%$ at the
wavelength 430 nm have been reported~\cite{imahori1}.

 Recently, a more efficient, sophisticated and rigid antenna-reaction system
has been designed in Refs.~\cite{gust2,gust3}. This system
includes three kinds of light-absorbing chromophores: (i) bis
(phenylethynyl)anthracene (BPEA), which absorbs around 450 nm
wavelength (blue region); (ii) borondipyrromethene (BDPY), having
a strong absorption at 513 nm (green region);  and (iii) zinc
tetraarylporphyrin, which absorbs both at 418 nm and at 598 nm.
This study reports $\sim$100$\%$ quantum yield for the excitation
transfer and $\sim$95 $\%$ quantum yield for the generation of the
charge-separated state P$^{+}$--C$_{60}^{-}$.

 Energy transfer mechanisms in the multi-chromophoric light-harvesting
complexes of bacterial photosystems (e.g., excitation-transfer
between chlorophyll molecules in the Fenna-Matthews-Olson (FMO)
protein complex) have been studied elsewhere (see, e.g.,
Ref.~\cite{fleming1,Gilmore,Alan,yang}). Those
works~\cite{fleming1,Gilmore,Alan,yang} have mainly focused on the
quantum effects in excitation-transfer across the
bacteriochlorophyll units.

Theoretical studies of antenna-reaction center complexes can be
useful for a better understanding of, and for optimizing
light-to-electricity conversion, as well as for developing new and
efficient designs of solar cells. Recently  we have theoretically
analyzed \cite{JPCc09} the light-to-electricity energy conversion
in a molecular triad (Fc--P--C$_{60}$) electronically coupled to
conducting leads. It was shown that the Fc--P--C$_{60}$ triad can
transform light energy into electricity with a power-conversion
efficiency of order of 40\%, provided that the connection of the
triad to the leads is strong enough. It should be noted, however,
that this prototype solar cell absorbs photons with energies in
close proximity to the resonant transition of the central
porphyrin molecule. Therefore, a major fraction of the sunlight
spectrum is not converted to the electrical form by this device.

In this paper we examine a theoretical model for the
light-to-electricity energy conversion by a molecular triad, which
is surrounded by four additional light-harvesting antenna
complexes. We show that this artificial photosystem is able to
generate a photocurrent with a quantum yield of the order of 90 \%
 (when the reorganization energy for the F\"orster transfer is
relatively high)  absorbing photons in a wide range (420--670 nm)
of the solar spectrum. We consider two different configurations
for the antenna complexes: (a) where each light-harvesting
molecule is independently connected (by the F\"orster
energy-transfer mechanism) to the central porphyrin (P) molecule
of the triad (Fig.~1a); and (b) where the light-harvesting
molecules are arranged in a line (Fig.~1b), with only one molecule
directly coupled to the porphyrin and with other molecules forming
a chain where the energy propagates in a cascade-manner.

Let us consider a (D-P-A) reaction center surrounded by several
(say, four) accessory light-harvesting antennas. Which is the best
way to arrange in space these accessory antennas? In other words,
which network or topology would provide more energy from the
surrounding accessory antennas to the central reaction center?  A
star-shaped configuration? (with each accessory antenna directly
coupled to the central reaction center).  Or in a somewhat
opposite configuration: as a linear chain with nearest-neighboring
couplings between the antennas and only one of these coupled to
the reaction center? These two cases are shown in Fig.~1.

In principle, very many possible topologies could be considered.
However, to simplify this analysis, we will now focus on two
extreme cases, with somewhat opposite topologies or networks: a
well-connected reaction center (directly connected to {\it all\/}
four accessory light harvesting antennas), and the opposite case
where the central reaction center is coupled to {\it only one\/}
antenna, which is now part of a linear chain. These two
extreme-opposite topologies or networks can be denoted as
star-shaped (shown in Fig.~1a) and linear-chain (Fig.~1b),
respectively.

At first sight, it would seem that the star-shaped configuration,
with each antenna directly connected to the reaction center, can
provide far more energy to the reaction center, due to the multiple
connections between the central unit, and the surrounding antennas.
However, an energy bottleneck can arise even in this
apparently-optimal topological configuration. Each antenna is here
assumed to operate optimally in a limited, and {\it shifted},
frequency range. In other words, we are not assuming all accessory
light-harvesting antennas to be equal to each other. In general,
various antennas can operate in different frequency ranges, and
this difference is crucial in our analysis. Recall that the
photosystem of green plants is made of six light-harvesting
accessory pigments: each pigment absorbing light in a
\emph{different range} of the solar spectrum.  Thus, in the
star-shaped topology shown in Fig.~1a, antennas with a large energy
mismatch would {\it not\/} provide energy to the central reaction
center. Only the surrounding accessory antennas that have an
approximate energy match to the central reaction center would
transfer energy. This energy bottleneck works against the
star-shaped network shown in (a).

Thus, the two issues considered here are: (1) how to physically
arrange antennas around the central reaction center, and also (2)
how to arrange these in {\it energy-space\/}, so to speak. The
first issue is topological and focuses on the network connectivity
in real space: for instance, how many accessory antennas are
connected to a central reaction center. The second issue refers to
the {\it energy match\/} (or mismatch) between neighboring
antennas, and between them and the central reaction center. A
large energy mismatch between any connected units in the chain
would preclude energy transfer between them. This approximate
``energy matching" issue between neighboring units is equally
important to keep in mind, not just the real-space topological
arrangement of the units.

As mentioned above, Fig.~1 shows the connectivity between the
different units: star-shaped topology in (a), and linear-chain
configuration in (b). Moreover, the colors  there represent, very
schematically, the energy range where each unit operates
optimally. The linear chain shown there has antennas arranged in a
way that nearby units operate in approximately similar energy
ranges.  This energy-matching issue perhaps is not very clear in
Fig.~1, even when seen in color. The energy scales are shown far
more explicitly in Fig.~2.  Figure 2b clearly shows that the
linear chain model considered here operates via an energy cascade,
or linear chain-reaction. Like a line of falling dominoes, one
event triggering the next one, in a sequential manner, with small
energy mismatches between successive energy transfer events.  As
shown in Fig.~2b, the more energetic antenna is located far away
from the reaction center, and it is coupled to an antenna with a
slightly lower energy, which is coupled to another antenna with an
even slightly lower energy, and so on, moving energetically
``downhill'' along the chain. Thus, the neighboring antennas must
be so both in real space, and {\it also\/} in ``energy space", to
allow for the efficient transfer of energy between them.  Thus,
``proximity'' between units must be in two spaces: real space and
energy space.

%%%%%%%%%%%%%%%%%%%%%   Figure 2   %%%%%%%%%%%%%%%%%%%%%%%%%%%%%%%%%%%%%%%%%%%

\begin{figure}[tp]
\centering
\includegraphics[width=9cm]{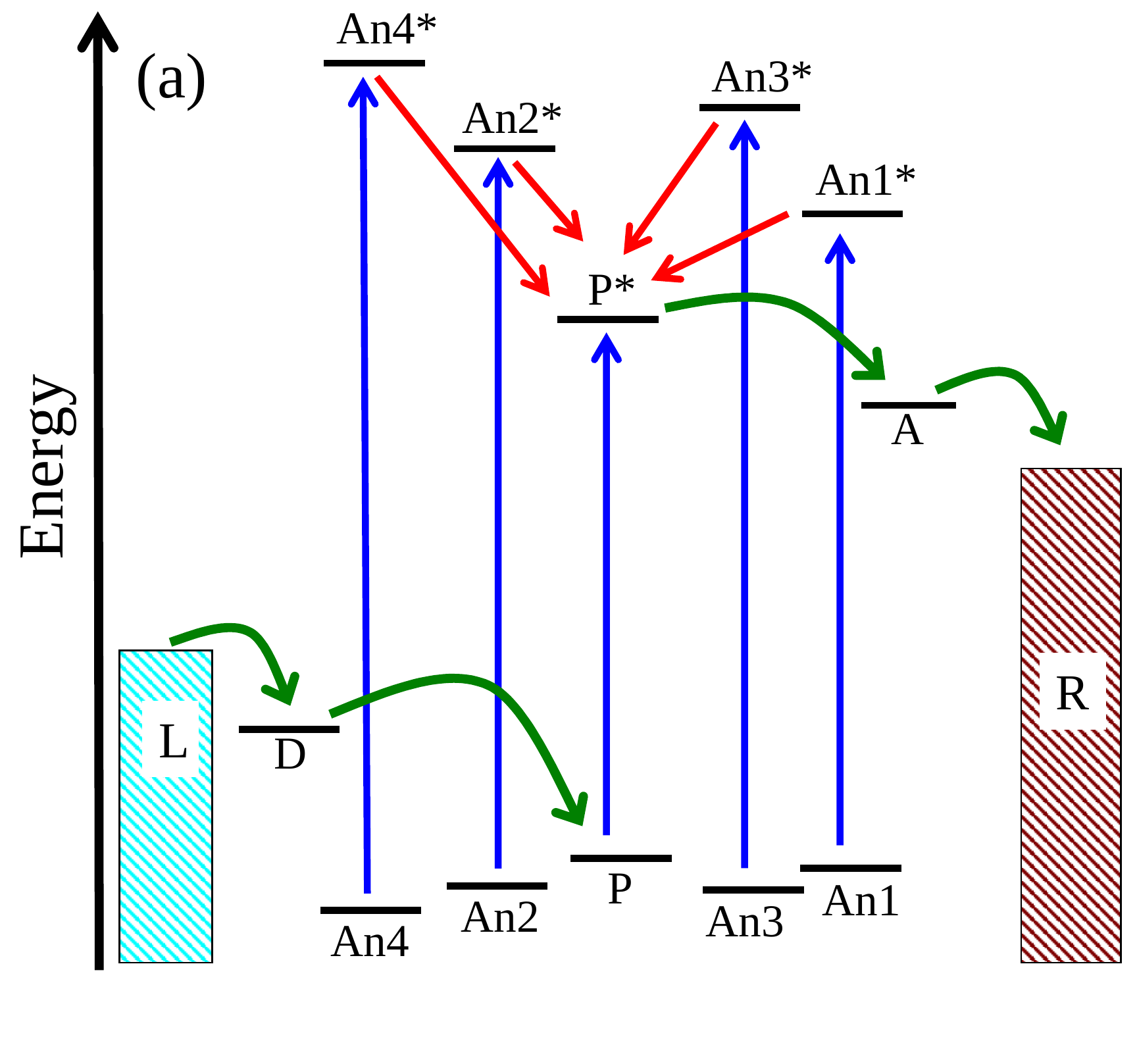}
\includegraphics[width=9cm]{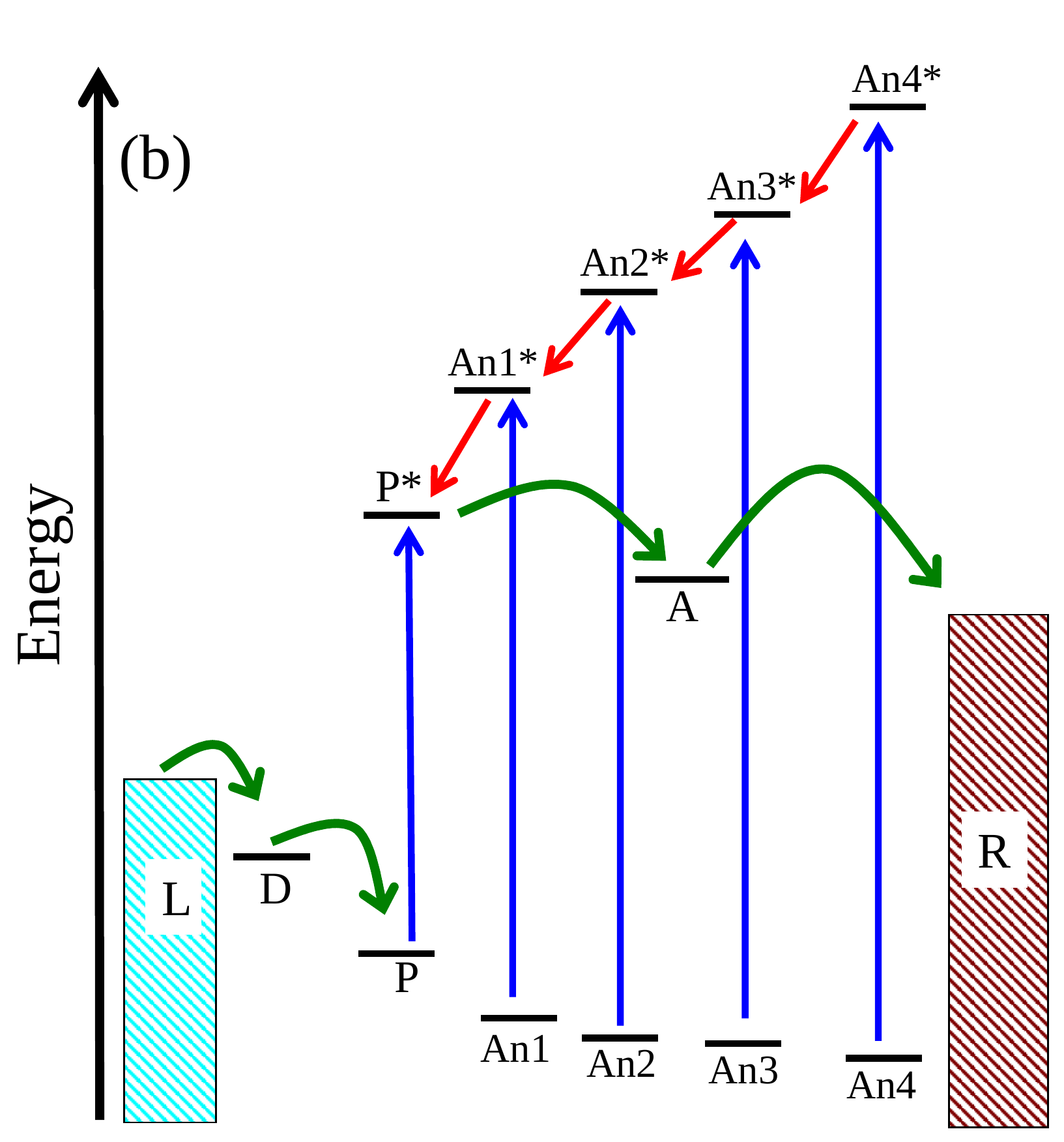}
\caption{(Color online) Energy diagram of the antenna complexes
An1, $\ldots$, An4, energetically coupled to the reaction center
(RC), D--P--A, for the case (a) where each light-harvesting
molecule is directly connected to the porphyrin molecule, P, of
the RC; and (b) where the energy transfer occurs via a
linear-chain of antenna with nearest-neighbor energy exchange.
Note that the four antenna chromophores in (b) are arranged in a
way to have a relatively small energy difference between
neighboring units.}
\end{figure}

%%%%%%%%%%%%%%%%%%%%%%%%%%%%%%%%%%%%%%%%%%%%%%%%%%%%%%%%%%%%%%%%%%%%%%%%%
This article is organized as follows: in  Section~II we outline a
model for the artificial reaction center (molecular triad) coupled
to the antenna complex.  We  briefly describe our mathematical
methods in Sec.~III.  The parameters are listed in Sec.~IV.  In
Sec.~V, we numerically solve the master equations and analyze the
energy transfer process. Conclusions are presented in Sec.~VI. The
methods used are described in more details in the online
supplementary material.

%\textbf{The paper is concluded in Sec.~IV. }

\section{Model}

We start with a schematic description of the energy and electron
transfer processes in an artificial reaction center D--P--A
(Donor--Porphyrin--Acceptor)  combined with four antenna
chromophores: An1, An2, An3, An4 (see Figs.~1a and 1b, showing two
different configurations for these antennas: star-shaped
configuration in 1a and chain in 1b). The photosensitive molecular
triad, D--P--A, is inserted between two electron reservoirs
(electrodes) L and R. The donor, D, is coupled to the left lead,
L, and the acceptor, A, is connected to the right lead, R.  As in
Refs.~\cite{imahori1,JPCc09}, the donor and acceptor molecules
(e.g., ferrocene and fullerene), are connected to each other via
the photosensitive molecule (porphyrin, P). This molecule is
surrounded by four Accessory Light-Harvesting Pigments (ALHP).
Figure~1a corresponds to the situation where all pigments are
directly coupled to the photosensitive part (P) of the molecular
triad. In this ``star-shaped" geometric arrangement, all the ALHP
can directly transfer excitations to the photosensitive part of
the molecular triad. Figure~1b corresponds to the case where the
light-harvesting pigments form a chain, which transfers energy
where the excitation moves  from one pigment to the next one via
 nearest-neighbor couplings: An4 $\rightarrow$ An3
$\rightarrow$ An2 $\rightarrow$ An1 $\rightarrow$ P. This
cascade-like excitation transfer occurs in an
energetically-downhill direction, akin a one-dimensional chain
reaction or domino-effect.

Figures~2a and 2b present the energy diagrams of the photosystems
described in Figs.~1a and 1b, respectively. The electron transfer
chain, L$\rightarrow$D$\rightarrow$P$\rightarrow$
P$^*\rightarrow$A$\rightarrow$R, is the same for both
configurations (a) and (b), and both begin on the left lead, L.
The electrochemical potentials of the left (L) and right (R)
electron reservoirs are determined by the parameters $\mu_L$ and
$\mu_R$, with $\mu_R > \mu_L.$ Since the energy level $E_D$ of the
donor D is lower than the potential $\mu_L$ of the left lead, $E_D
< \mu_L$, electrons can move from the L-reservoir to the level D
and, afterwards, to the low-lying ground energy level, $E_P$, of
the porphyrin. When absorbing a photon, the electron in the
porphyrin molecule jumps from its ground state P to its excited
state P$^*$. A subsequent electron transfer from P$^*$ to the
acceptor A is driven by a negative energy gradient, $(E_A -
E_{P^*}) < 0$.  In view of the relation: $E_A > \mu_R,$ the
electron in A is finally transferred to the right, R, electron
reservoir. This is the light-induced electron transition in the
porphyrin molecule, which results in an energetically-uphill
electron flow in both photosynthetic systems (a) and (b).

Even though these systems (a) and (b) have the similar electron-transport chains, their light-harvesting complexes are arranged quite
differently. Each of these complexes, An = An1, $\ldots$, An4, can be characterized by a ground, $E_{An}$, and an excited, $E_{An^{*}},$ energy
levels with an energy difference $\omega_{An} = E_{An^{*}} - E_{An}$. Hereafter, we assume that $\hbar =1$ and $k_B = 1$. For the
light-harvesting complex (a) (see Figs.~1a, 2a) all frequencies $\omega_{An1},\ldots,\omega_{An4}$ should exceed the porphyrin transition
frequency, $\omega_P = E_{P^{*}} - E_P.$  In this case the energy of photons collected by each individual antenna can be transferred directly to
the photosensitive part of the artificial reaction center (porphyrin molecule). However, in the light-harvesting complex (b) (see Figs.~1b and
2b) only the antenna An1 is coupled (by a F\"orster mechanism) to the porphyrin, whereas the other light-harvesting pigments form a linear chain
that transfers energy downhill along the chain: An4 $\rightarrow$ An3 $\rightarrow$ An2 $\rightarrow$ An1 $\rightarrow$ P. This energy transfer
can be energetically-allowed provided that $\omega_{An4} > \omega_{An3} > \omega_{An2} >\omega_{An1} > \omega_{P}.$ In Sec.~V we compare these
two artificial photosystems and determine which arrangement of the antenna complexes provides more energy to the reaction center.

\section{Methods}

The electron flow through a molecular triad coupled to two
electron reservoirs can be described with methods of quantum
transport theory and the theory of open quantum systems
\cite{Wingr93,MarcusSutin,JCP09CcO,JCP09Photo}. In addition to
four sites (D, P, P$^*$, A), describing the molecular triad, we
introduce four pairs (An1, An1$^*$, $\ldots$, An4, An4$^*$), which
characterize the ground and excited states of the light-harvesting
antennas. The whole system, the molecular triad plus four antenna
complexes, can be analyzed within a mathematical formalism
presented in Supporting Information and also in
Ref.~\cite{JPCc09}.

The total Hamiltonian of the system includes the following
components:

\noindent (i) energies of the electron sites and leads, as well as
the Coulomb interactions between the electrons located on
different sites of the triad;

\noindent (ii) tunneling couplings between the electron sites on
the triad and the electron reservoirs;

\noindent (iii) electron tunneling between the electron sites
belonging to the molecular
 triad;

 \noindent (iv) coupling of electron sites and antenna complexes to an environment;

 \noindent (v) interactions of the porphyrin molecule and antenna complexes
  with an external electromagnetic
 field (laser field) and with

 \noindent (vi) blackbody radiation and Ohmic bath, responsible
  for the quenching (energy loss) of the porphyrin and antenna excited
 states.

 \noindent (vii) In the case of the design in Fig.~1a the Hamiltonian
includes the direct F\"orster coupling between the porphyrin
molecule, P, and the light-harvesting complexes An1, $\ldots$,
An4. For the linear-chain configuration shown in Fig.~1b, the
F\"orster mechanism provides the energy transfer between the
nearest-neighbors in the antenna chain, as well as between the
complex An1 and the porphyrin molecule.

\section{Parameters} \emph{Energy levels and electrochemical
potentials:} The energy levels of the Fc--P--C$_{60}$ molecular
triad are $E_D=-510$ meV, $E_P =-1150$ meV, $E_{P^*}= 750$ meV and
$E_A = 620$ meV. These values are obtained by estimating the
reduction potentials (using a reference electrode Ag/AgCl) of
ferrocene (D), porphyrin (P, P$^*$) and fullerene (A) molecules
\cite{JPCB2000}. For the electrochemical potentials of the left
($\mu_L$) and the right ($\mu_R$) leads, we choose the following
values: $\mu_L = -410$~meV and $\mu_R = 520$~meV, with the
electrochemical gradient $\Delta \mu = \mu_R - \mu_L = 930$~meV.

\emph{Coulomb interactions:} The spatial separations between D--P,
P--A and D--A are of order of 1.62 nm, 1.8 nm and 3.42 nm,
respectively \cite{JPCB2000}. The Coulomb energies $u_{DP}$,
$u_{DA}$ and $u_{PA}$ can be calculated with the formula
\begin{eqnarray}
u_{ij}=\frac{e^2}{4\pi\epsilon_0 \epsilon r_{ij}} \nonumber
\end{eqnarray}
where, $\{ij\} \;= {\rm \;\{DP\},\;\{DA\},\;\{PA\},}$ and
$\epsilon_0$ is the vacuum dielectric constant. For $\epsilon \sim
4.4$, the Coulomb interaction energies are $u_{DP}= 200$ meV,
$u_{DA} = 95$ meV and $u_{PA} = 180$ meV.

The F\"orster coupling, $V_F$, between the photosensitive
molecules $l$ and $l'$ is proportional to the product of the
dipole moments of these molecules, $e r_{l}$ and $e r_{l'}$, and
inversely proportional to the cubic power of the distance
($R_{ll'}$) between them \cite{PumpPRE08}:
\begin{equation}
V_F =\frac{e^2}{2\pi \epsilon_0 \epsilon}\frac{r_{l}r_{l'}}{R^3}
\end{equation}
For the case where $r_k \sim r_l \sim$ 0.3 nm, $R \sim $1 nm, and at
$\epsilon \sim$ 4,  the F\"orster coupling is about $V_F \sim$
65~meV.

\emph{Tunneling amplitudes:} We have assumed the
ferrocene--porphyrin and porphyrin--fullerene tunneling amplitudes
are about $\sim$ 3 meV, so that
$$
\frac{\Delta_{DP}}{\hbar} =\frac{\Delta_{DP^*}}{\hbar} =
\frac{\Delta_{AP}}{\hbar} = \frac{\Delta_{AP^*}}{\hbar} = 4.5 \;
\rm{ps}^{-1}.
$$

For the tunneling rates between the left lead and ferrocene
($\Gamma_L$) and between the right lead and fullerene ($\Gamma_R$)
we choose the following values \cite{JPCc09}: $\Gamma_L/\hbar$ =
1800 $\mu$s$^{-1}$ and $\Gamma_R/\hbar$ = 180 $\mu$s$^{-1}$.

\emph{Radiation leakage and quenching rates:} We take the
following estimates for the radiation leakage time: $\tau_{P^* \to
P} = \tau_{P^* \to D} = \tau_{A \to P} \sim 0.4$ ns. Similar
estimates have been used for the radiation leakage timescales of
the antenna molecules. For the quenching (or energy-loss) time of
the porphyrin excited state P$^*$ we use the value:
$\tau_{\rm{quen}} \sim 0.1$~ns.

\emph{Reorganization energies for the electron and energy transfers:} For  the molecular triad analyzed in Ref.~\cite{JPCc09} we obtain the
relatively high power-conversion efficiency, $\eta \sim 42 \%,$ provided that the donor-porphyrin and acceptor-porphyrin electron transfer
reorganization energies are about $\Lambda_{DP} \sim 600$~meV and $\Lambda_{AP} \sim 100 - 400$~meV. A much smaller value, $\Lambda_{PP^*} \sim
100$~meV, is assumed for the light-induced electron transitions between the ground (P) and excited (P$^*$) levels of the porphyrin molecule. The
F\"orster energy transfer between the light-harvesting molecules and between these molecules and the porphyrin is also accompanied by an
environment-reorganization process, which can be characterized by a smaller energy scale, $\Lambda_F \leq 100$~meV (see, e.g., the energy
transfer in the B850 complex \cite{B850}, where $\Lambda_F$ is assumed to be about 30~meV).

Hereafter, we assume that the external light source has a fixed
intensity, $I = 100$~mW/cm$^2$, and that the environment is kept at
the room temperature, $T = 298$~K. We also assume that the
reorganization energy for the F\"orster energy transfer,
$\Lambda_F,$ is about 100~meV, unless otherwise specified.

We analyze the Fc--P--C$_{60}$ molecular triad, where the
ferrocene molecule, Fc, is attached to the gold surface (left
lead, L), and the fullerene, C$_{60}$, is in contact with an
electrolyte solution (right lead, R) filled with oxygen molecules,
which are able to accept electrons from the C$_{60}$ molecules.

\section{Results and Discussion}

We derive and  solve numerically a set of master equations for the
probabilities to find the system in a definite eigenstate of the
basis Hamiltonian. This is explained in the online supplementary
material. After that, we calculate: (i) the energetically-uphill
electron current through the triad, (ii) the energy of the photons
absorbed by the triad and by the light-harvesting molecules. This
allows us to determine a quantum yield and power-conversion
efficiency of the system (see all definitions in the supporting
information).

\subsection{Photocurrent through the molecular triad
directly coupled to four porphyrin light-harvesting molecules}

Here we consider the situation where both the reaction center and
the antenna complexes are made of porphyrin molecules with the
geometrical arrangement shown in Fig.~1a. This arrangement allows
direct energy transfer from each light-harvesting chromophore to
the reaction center.

%%%%%%%%%%%%%%%%%%%%%%%%%   Figure 3 %%%%%%%%%%%%%%%%%%%%%%%%%%%%%%%%%%%%%%%%

\begin{figure}[tp]
\centering
\includegraphics[width=8cm]{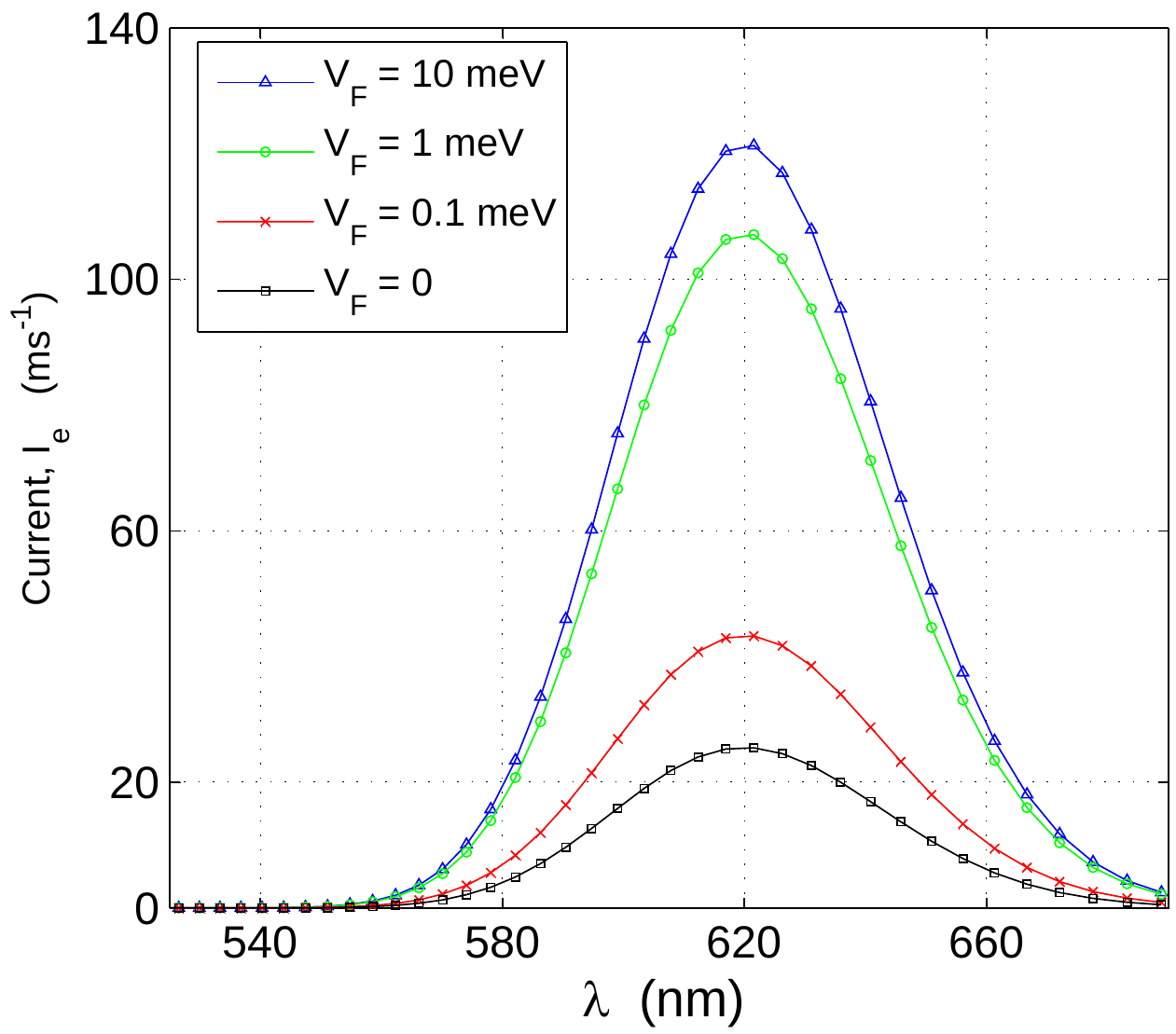}
\caption{(Color online) Photoinduced electron current $I_e$ (
number of electrons, in one ms, pumped from the L to the R lead)
versus the wavelength of the incident light for a photosystem with
four antenna complexes, which are made of porphyrin molecules.
Both the antenna complexes and the reaction center absorb at the
same wavelength. The whole complex now absorbs more photons than
the single porphyrin molecule; thus, pumping many more electrons
from the left (with $\mu_L = - 410$~meV) to the right (with $\mu_R
= 520$~meV) electron reservoir. The other parameters are listed in
the text (see Sec. IV). The peak in the current increases for
larger values of the  F\"{o}rster coupling strength $V_F$. We also
studied (not shown here) higher values of $V_F$, but these
produced the same results as $V_F = 10$ meV. Thus, this value of
$V_F$ provides a saturation in the electron current. The resonant
peak here is $\lambda = 620$ nm. }
\end{figure}

%%%%%%%%%%%%%%%%%%%%%%%%%%%%%%%%%%%%%%%%%%%%%%%%%%%%%%%%%%%%%%%%%%

In Fig.~3 we plot the photocurrent through  the triad as a
function of the wavelength of light for different values of the
F\"{o}rster coupling strength : $V_F = 0, 0.1, 1, 10$ (in meV) and
for the above-mentioned set of parameters of the system.  It is
apparent from Fig.~3 that the magnitude of the light-induced
pumping current at $\lambda = 620$~nm is significantly enhanced
(about 5 times larger when $V_F = 10$~meV) by the antenna system.
However, the spectral range of the light absorption remains the
same as for the detached porphyrin reaction center (see Fig.~3,
where the red curve with square symbols describes the photocurrent
through the molecular triad completely disconnected from the
antenna chromophores, $V_F = 0$). We also find that the quantum
yield, $\Phi$, taken in the middle of the resonant peak ($\lambda
= 620$~nm), \emph{non-monotonically} depends on the F\"orster
coupling strength $V_F$ measured here in meV: $ \Phi(V_F = 0)
\simeq 0.85; \; \Phi(0.1) \simeq 0.3; \; \Phi(1) \simeq 0.75$ and
$ \Phi(10) \simeq 0.82.$

\subsection{Molecular triad connected to two BPEA and two BDPY chromophores }

Now we consider a different case: an antenna system comprised of
two BDPY and two BPEA molecules. The BDPY molecule has the maximum
absorbance in the green region (at 513 nm) of the solar spectrum,
where neither BPEA (with maximum absorbance at 450 nm) nor the
porphyrin, which absorbs at 620 nm, have maxima of absorption
spectra. It should be noted that a multichromophoric hexad antenna
system having three light-absorbing BDPY, BPEA and porphyrin has
been developed in Ref.~\cite{gust2}. The generation of the
charge-separated state, P$^{+}$--C$_{60}^{-}$, with almost 95$\%$
quantum yield~\cite{gust2}. In our case, the porphyrin unit of the
molecular triad is coupled to the four antenna chromophores (two
BDPY and two BPEA).

%%%%%%%%%%%%%%%%%%%%%%%%%%%%%    Figure 4 %%%%%%%%%%%%%%%%%%%%%%%%%%%%%%%%%%%%%%%%%%%%

\begin{figure}[tp]
\centering
\includegraphics[width=8cm]{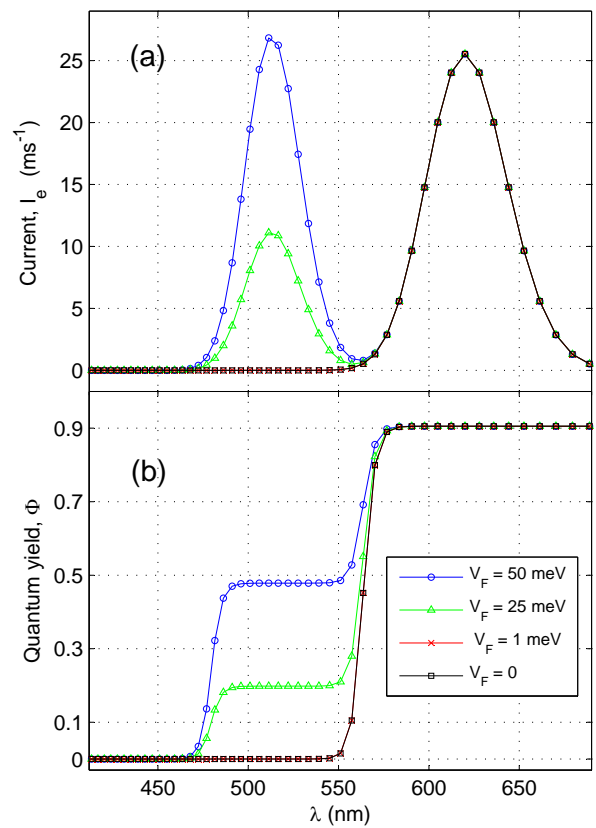}
\caption{(Color online) Electron current $I_e$ and  quantum yield
$\Phi$ as functions of the wavelength $\lambda$ of the external
radiation for the configuration shown in Fig.~1a, where two BPEA
and two BDPY antenna chromophores are directly coupled to the
centrally-located reaction center. The F\"orster coupling
constant, $V_F$, which is assumed to be the same for every
chromophore-RC connection, takes four values (in meV): $V_F =
0,\;1,\;25,\;50$. For other parameters see Sec. IV. Note that the
electron current and the quantum yield grow for increasing values
of the F\"orster coupling energy strength $V_F$. More importantly,
the direct coupling (Fig.~1a) suppresses the peak at $\lambda$=
450 nm (Fig.~4a), which is present in the linear chain
configuration (Fig.~1b), as shown in Fig.~5a }
\end{figure}

%%%%%%%%%%%%%%%%%%%%%%%%%%%%%%%%%%%%%%%%%%%%%%%%%%%%%%%%%%%%%%%%%%%%%%%%%

We consider two situations: (a) where the antenna chromophores are
\emph{directly} coupled to the porphyrin unit of the reaction
center (Fig.~1a); and (b) where the antenna chromophores are
\emph{arranged in line}: BDPY $\rightarrow$ BDPY $\rightarrow$
BPEA $\rightarrow$ BPEA $\rightarrow$ RC, with the
nearest-neighbor coupling between chromophores. See Fig.~1b. Thus,
the configuration (a) might appear to be energetically more
efficient than (b). However, our calculations below indicate that
this is not the case.
%%%%%%%%%%%%%%%%%%%%%   Figure 5 %%%%%%%%%%%%%%%%%%%%%%%%
\begin{figure}[tp]
\centering
\includegraphics[width=8cm]{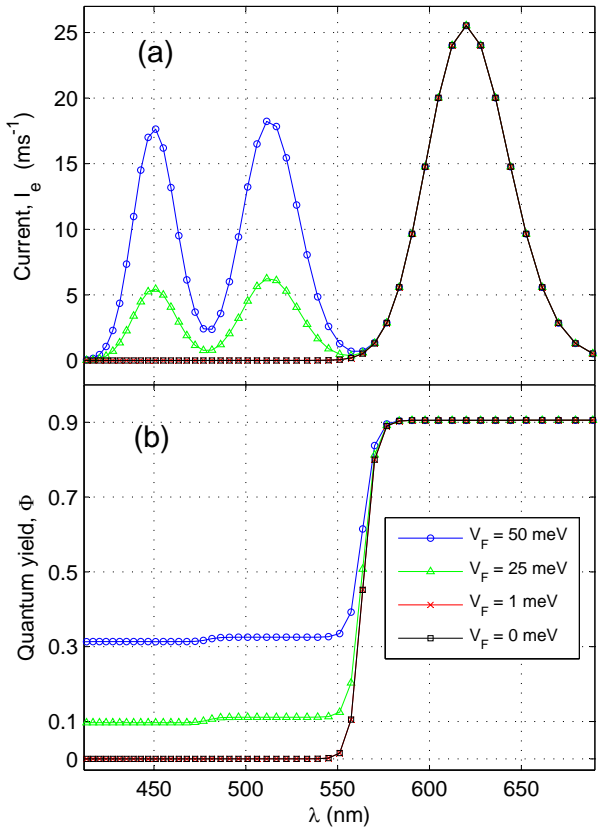}
\caption{(Color online) Electron current $I_e$ and a quantum yield
$\Phi$ versus the wavelength of light, $\lambda,$ for the
configuration shown in Fig.~1b, where the excitation energy moves
along the following chain of light-harvesting molecules: BPEA
$\rightarrow$ BPEA $\rightarrow$ BDPY $\rightarrow$ BDPY
$\rightarrow$ RC. The parameters used here are the same as in
Fig.~4. However, the current peak at $\lambda \sim$ 450 nm is
present in (a) here, but absent in Fig.~4(a), which used a
direct-coupling configuration to the central reaction center. }
\end{figure}

\noindent (a) For the case of direct connection between the four
antennas chromophores and the triad (see Fig.~1a and Fig.~4) we
calculate a photocurrent and a quantum yield, $\Phi$, as functions
of the wavelength of light, at $\Lambda_F = 100$~meV, and at five
values of the F\"orster coupling: $V_F = 0,\;1,\;25,\;50$~meV. The
wavelength dependence of the current has two maxima centered at
513 nm and 620 nm. The BPEA molecules, which absorb at 450~nm,
give a negligible contribution to the current since their spectral
maxima are too far from the absorbance maximum of the porphyrin
spectrum. As a consequence, the BPEA-porphyrin energy transfer is
significantly suppressed at moderate values of the F\"orster
reorganization energy, $\Lambda_F \leq 100$~meV in the range of
the coupling constants $V_F \leq 50$~meV. It follows from Fig.~4
(see a peak at $\lambda = 513$~nm) that the BDPY molecules start
working as efficient light-harvesters only at sufficiently strong
F\"orster coupling, $V_F \geq 10$~meV, to the porphyrin unit of
the molecular triad. We also note that when $\lambda \sim 513$~nm,
both the photoinduced current and the quantum yield grow with
increasing the F\"orster coupling strength, so that the quantum
yield, $\Phi$, can be around  48\%. In the range of porphyrin
absorption (at $\lambda \sim 620$~nm and $V_F = 0$) the quantum
yield is  of the order of 90$\%$.
%%%%%%%%%%%%%%%%%    Figure 6        %%%%%%%%%%%%%%
\begin{figure}[tp]
\centering
\includegraphics[width=8cm]{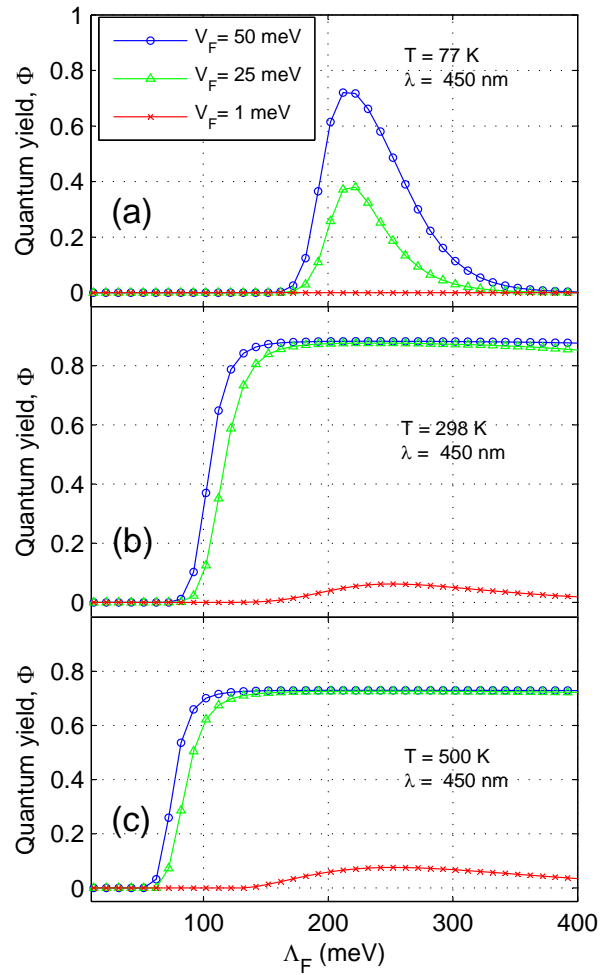}
\caption{(Color online) Quantum yield as a function of
reorganization energy, $\Lambda_F$, for the linear-chain
nearest-neighbor coupling between chromophores, BPEA $\rightarrow$
BPEA $\rightarrow$ BDPY $\rightarrow$ BDPY $\rightarrow$ RC, for a
the wavelength of light $\lambda = 450$~nm (blue peak in Fig.~5a).
This figure is plotted for three different temperatures: (a) Very
low $T$ = 77~K, (b) room temperature  $T$ = 298~K and (c) very
high $T$ = 500~K as well as for three values of the F\"orster
constant $V_F = 1,\;25,$ and 50~meV. As shown in the figures (b)
and (c), increasing the reorganization energy $\Lambda_F$, can
sharply increase the quantum yield. The extreme low-temperature
case in (a) is just a limit case, shown for comparison with the
higher-temperature cases in (b) and (c).}
\end{figure}

\noindent (b) A much broader light spectrum can be converted into
electrical current in the linear configuration in Fig.~1b, where
the light-harvesting chromophores  are arranged along a line: BPEA
$\rightarrow$ BPEA $\rightarrow$ BDPY $\rightarrow$ BDPY
$\rightarrow$ RC, with the only one BDPY molecule directly coupled
to the porphyrin unit of our artificial reaction center (RC) (see
Fig.~1b and Fig.~5). This system is able to collect photons in the
range of wavelength from 420~nm up to 650~nm covering a
significant part of the visible sunlight spectrum.  The chain of
BPEA and BDPY molecules creates an efficient channel, which
gradually transmits energy from the collectors of high-energy
photons (BPEA molecules), via the intermediate BDPY antennas, to
the molecular triad. In Figs.~5a, 5b we plot  the photoinduced
current and the quantum yield versus the wavelength of the
external radiation, at $\Lambda_F = 100$~meV, and at four values
of the F\"orster constants (in meV) $V_F = 0, \;1,\;25,\;50.$ For
large value of the F\"{o}rster coupling strength $V_F$ the quantum
yield $\Phi$ reaches $\sim$ 48\% in Fig.4(a) around $\lambda =$
513~nm [for the star-shaped topology in Fig.~1a] and $\sim$ 30\%
in Fig.~5(b) [for the linear chain case in Fig~1b]. The quantum
yield in Fig.~5b is lower than the one in Fig.~4a, but extends
over a wider range of wavelengths including the peaks around
$\lambda \sim$ 450~nm and $\lambda \sim$ 513~nm.

%%%%%%%%%%%%%% Figure 7 %%%%%%%%%%%%%%%%%%%%%%%%%%%%%%%%%%

\begin{figure}[tp]
\centering
\includegraphics[width=8.7cm]{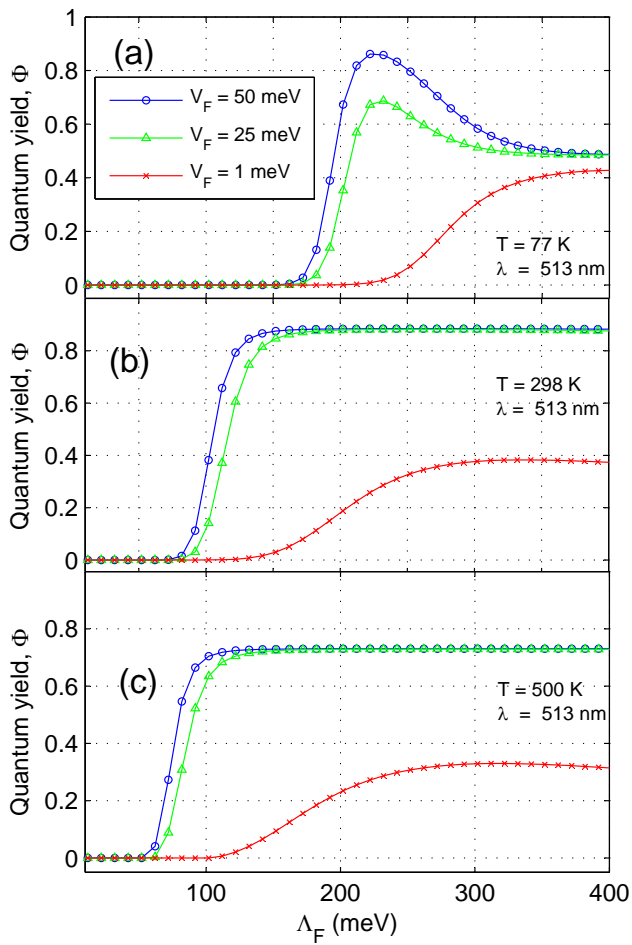}
\caption{(Color online) Quantum yield versus reorganization
energy, $\Lambda_F$, for the green peak ($\lambda = 513$~nm) of
the spectrum (see Fig.~5a) for three values of the F\"orster
constant, $V_F = 1,\;25,\;50$~meV, and at three different
temperatures: (a) $T$ = 77~K, (b) $T$ = 298~K and (c) $T$ = 500~K.
The other parameters are listed in Sec.~IV. Figures 5, 6, and 7
focus on the linear chain configuration (Fig.~1b) with
nearest-neighbor couplings between chromophores. Figures 6 and 7
show the same quantities, but centered at different peaks
($\lambda \sim $ 450 nm versus  $\lambda \sim$ 513 nm).}
\end{figure}

It follows from Fig.~4 and Fig.~5, that the configuration using
the chain-like nearest-neighbor coupling between light-harvesting
chromophores [Fig.~1b, design (b)] converts much more blue
($\lambda = 450$~nm) light into electricity with a higher quantum
yield than the star-shaped configuration with  direct coupling
between the antenna chromophores and the RC [Fig.~1a, design (a)].
In this star-shaped configuration,  the BPEA molecules, which
collects blue photons, are \emph{not} able to transfer their
energy to the photosensitive unit of the triad due to a
significant difference between the energies of the BPEA antennas
($\lambda = 450$~nm) and the porphyrin-based reaction center
($\lambda = 620$~nm).

The antenna-RC energy transfer is facilitated by the strong
coupling to the environment (when the energy difference is high),
characterized by the reorganization energy $\Lambda_F$, as well as
by a tight F\"orster
 binding between chromophores, which is
 described by the constant $V_F$. In Figs.~6 and 7  we plot
the quantum yield, $\Phi$, as a function of the reorganization
energy, $\Lambda_F$, for three different temperatures (in K): $T$
= 77, 298, 500, and  for three values
 of the coupling constant $V_F = 1,\;25,\;$ and 50~meV.
Figure~6 is related to the blue peak of the absorption spectrum
($\lambda = 450$~nm), whereas figures 7 describes the behavior of
the green peak ($\lambda = 513$~nm). The peak centered at $\lambda
= 620$~nm is produced by the porphyrin molecule, belonging to the
triad, and, therefore, shows no dependence on $\Lambda_F$ and
$V_F$.

The Marcus rates, describing the energy transmission  between the
photosensitive elements of the system, depend  (i) on the energy
difference, $\Delta E$, between the photosensitive units, (ii) on
the F\"{o}rster coupling $V_F$, and (iii) on the F\"orster
reorganization energy $\Lambda_F$. The large energy separation,
$\Delta E$, of the energies of the nearby photosensitive molecules
leads to a decrease of the Marcus rates and, thus, to the
suppression of the energy transfer. In our case, the energy
distance between the BPEA ($\lambda$ = 450 nm) and BDPY ($\lambda$
= 513 nm) molecules is about 340~meV, whereas the energy
separation of the BDPY chromophore and porphyrin ($\lambda$ = 620
nm) is of order of 415~meV. This energy gap can be partially
compensated by the large reorganization energy, $\Lambda_F$, which
reflects the significant fluctuations of the relative positions of
the energy levels. Here, the environment plays a positive role
assisting the efficient and fast energy transfer between
chromophores (see also Ref.~\cite{Plenio,Aspuru08}). The
timescales for the energy and electron transfers should be shorter
than the radiation leakage time and the quenching time, otherwise
the energy of the photons absorbed by the system will be lost.

At very low temperatures (e.g., liquid nitrogen, $T$ = 77 K), the
fluctuations in the positions of the energy levels of the
chromophores are frozen, so that the light-to-electricity
conversion requires sufficiently large values of the
reorganization energy, $\Lambda_F
> 180$~meV (see Figs.~6a and 7a). Note that for $V_F
> 25$~meV the linear-chain arrangement system has an optimal performance at $\Lambda_F
\simeq 225$~meV for both frequency ranges. This extreme low
temperature case is only shown for comparison with the higher
temperatures cases.

At room temperature ($T$ = 298 K) and at strong enough F\"orster
coupling, $V_F \geq 25$~meV,  the  blue and green spectral peaks
demonstrate similar behaviors as functions of $\Lambda_F$ (see
Figs. 6b and 7b). Here, the quantum yield begins to grow when the
reorganization energy exceeds $\sim 80$ meV, reaching finally 90\%
at $\Lambda_F > 150$~meV. These numbers are determined by the
parameters of the antenna-triad complex, and, especially, by the
radiation leakage time $\tau_{{\rm rad}}$, of the excited
porphyrin state P$^*$, estimated above as $\tau_{{\rm rad}} \sim
0.4$~ns.

At high temperatures (see Figs.~6c and 7c, plotted for $T$ =
500~K) the facilitating effect of the environment increases, and
the efficient energy transfer starts at the lower reorganization
energies, $ \Lambda_F \geq 75$~meV. This value of $\Lambda_F$ is
comparable with the reorganization energy for the energy transfer
in the B850 light-harvesting complex, where $\Lambda_F \simeq
27$~meV \cite{B850}.

At the smaller value of the F\"orster coupling, $V_F = 1$~meV, the
conversion of the blue light ($\lambda$ = 450 nm) to electricity
is significantly suppressed (see Fig.~6), whereas for green light
($\lambda$ = 513 nm) the dependence of the quantum yield on
$\Lambda_F$ are shifted to higher reorganization energies
(Fig.~7), compared to the case of the larger couplings, $V_F =
25,\; 50$~meV.

It should be noted that, to cover a broader range of the spectrum
of light with a fixed  number of antenna chromophores, the
resonance energies of the light-harvesting complexes should be
very well separated. However, in this case the energy transfer
between the antenna chromophores would be quite slow, since this
transfer is governed by the rates corresponding to the inverted
regions of the Marcus parabola. Then, the dissipation comes into
play, and the energy of the absorbed photons is lost on its way
from the antennas to the reaction center. The energy transfer
rates and, thus, the efficiency of the system can be maximized in
the case when the energy distance, $\Delta E_{i} = E_{i+1} -
E_{i},$ between the nearby light-harvesting complexes (labeled by
indices $i+1$ and $i$, with energies  $E_{i+1}$ and $E_{i}$) is
equal to the corresponding reorganization energy, $\Lambda_F^{i}$.
That is, $\Delta E_{i} = E_{i+1} - E_{i} = \Lambda_F^{i}.$

\section{Conclusions}

In this paper we have studied theoretical  aspects of the operation
of an artificial reaction center (a ferrocene--porphyrin--fullerene
molecular triad) coupled to the complex of four light-harvesting
molecules. We have analyzed two configurations of the antenna
complex: (a) a star-shaped configuration, where each
light-harvesting molecule is able to transfer energy directly to
the centrally-located reaction center, and (b) a case where the
antenna molecules form a linear chain, which gradually transfers
excitations from the high-energy antenna located in the far end, to
the antenna chromophore with the lowest energy. The last antenna
chromophore in the chain is energetically connected to the reaction
center (RC). To be specific, we have considered the case when the
antenna complex is comprised of two molecules of
bis(phenylethynyl)anthracene (BPEA), absorbing blue photons
($\lambda$ = 450 nm), and two molecules of borondipyrromethene
(BDPY), having an absorption maximum in the green region ($\lambda$
= 513 nm). We have shown that the configuration with a linear
arrangement of the antenna chromophores (configuration (b)) is able
to convert blue and green photons to electricity with a quantum
yield of order of $\sim$30\% (over a wide range of wavelengths),
whereas the energy of the red photons, absorbed by the molecular
triad itself ($\lambda$ = 620 nm), is converted to a current with a
quantum yield reaching the value of 90\%. We have investigated
dependencies of the quantum yield on the F\"orster reorganization
energy as well as on the F\"orster coupling constants between
chromophores and have shown that the environment plays a
significant role in facilitating the antenna-RC energy transfer,
thus, improving the light-harvesting function of the system.
Overall, the configuration (b) is more efficient than (a) in
transferring energy to the reaction center.

We emphasize that the artificial photosystem analyzed in this work
can be implemented with real light-harvesting components, such as
porphyrin and BPEA/BDPY molecules. The excitonic (F\"orster)
coupling strongly depends on the mutual distances and the
orientations of the chromophores. Similar to the wheel-shaped
antenna-reaction center complex implemented in Ref.~\cite{gust2},
the components of the photosystem can be placed at distances of the
order 10 \AA, which allows for a sufficiently strong F\"orster
coupling between the antenna chromophores and the reaction center.
At the same time, the chromophores comprising the light-harvesting
complex retain their individual molecular features. The
reorganization energy, another controlling parameter for energy
transfer, is varied for the system under study. Namely, we
numerically calculate both the light-induced electron current and
the quantum yield as functions of the reorganization energy. This
allows us to determine the value of the reorganization energy at
which the system works with maximum optimal efficiency.

\textbf{Acknowledgements.} FN acknowledges partial support from the
Laboratory of Physical Sciences, National Security Agency, Army
Research Office, DARPA, National Science Foundation grant No.
0726909, JSPS-RFBR contract No. 09-02-92114, Grant-in-Aid for
Scientific Research (S), MEXT Kakenhi on Quantum Cybernetics, and
Funding Program for Innovative Research and Development  on Science
and Technology (FIRST).

\appendix
\begin{appendix}
\section{Hamiltonian} Here we describe the methods used in our work.
We characterize the electrons in the states $i$ (= D, P, P$^*$,
An1, An1$^*$, An2,  An2$^*$, An3, An3$^*$, An4, An4$^*$, A) by the
Fermi operators $a_i^+$ and $a_i$ with the electron population
operator $n_i = a_i^+ a_i$. Each electron state can be occupied by
a single electron, as the spin degrees of freedom are neglected.
Electrons in the leads (electrodes) are described by the Fermi
operators $d_{k\alpha}^+,d_{k,\alpha}$, where $\alpha = \rm {L,R}
$; and $k$ is an additional parameter which has the meaning of a
wave vector in condensed matter physics. The number of electrons in
the leads is determined by the operator $\sum _{k}N_{k\alpha}$,
with $N_{k\alpha}=d_{k\alpha}^+ d_{k\alpha}$. The total Hamiltonian
of the system is  complicated. It includes the terms described
below.

\subsection{Eigenenergies and Coulomb interactions}
This part of the Hamiltonian involved the eigenenergies of the
electron states ($i$ = D, P, P$^*$, An1, An1$^*$, An2,  An2$^*$,
An3, An3$^*$, An4, An4$^*$, A) and the Coulomb interactions between
the electron states.
\begin{eqnarray}
H_{0} &=& \sum _{i} E_i n_i  + u_{P}n_P n_{P^*} + u_{DP} (1-n_{
D})\left(1-n_{P}-\right. \nonumber\\
&&\left.n_{P^*}\right)
- u_ {DA} (1-n_ {D})n_{A} - u_{PA}(1-n_{P}-n_{P*})n_{A}.\nonumber\\
\end{eqnarray}
 The symbols $u_{P}, u_{DP}, u_{DA}, u_{PA}$ represent
the electrostatic interactions between the electron  sites. We have
 assumed that the empty donor state D (with $n_{D}=0$)
as well as the empty photosensitive group ($n_{P} + n_{P^*} = 0$)
have positive charges. Therefore, $U_{DP} > 0$ because both D and P
are positively charged and thus repulsive.   The acceptor state $A$
becomes negatively charged when it is occupied by an electron and
thus $- U_{DA} < 0$ and $- U_{PA} < 0$. This attraction occurs when
the acceptor A is occupied ($n_A = 1$) and the D and P states are
both empty.  Also,  the acceptor state A is neutral when it is
empty.

\subsection{F\"{o}rster couplings}
We consider the energy transfer between the reaction center (P) and
the antenna complexes, and also among the antenna complexes by
introducing F\"{o}rster coupling terms,
\begin{eqnarray}
H_{{\rm Forster}} =
-\sum_{kl}V_{kl}\,a_{l}^{+}\,a_{l^*}\,a_{k^*}^{+}\,a_{k} + {\rm
H.c.}\;,
\end{eqnarray}
where, the pair $\{ k,l\}$ = \{P,~An1\}, \{P,~An2\}, \{P,~An3\},
\{P,~An4\}, \{An1,~An2\}, \{An2,~An3\}, \{An3,~An4\}. Here, $V_F$
determines the strength of the F\"{o}rster coupling and is
proportional to the product of the dipole moments, $e r_{l}$ and $e
r_{k}$, and inversely proportional to the cubic power of the
separating distance, $R$, between the chromophores,
\begin{eqnarray}
V_F =\frac{e^2}{2\pi \epsilon_0 \epsilon}\frac{r_{k}r_{l}}{R^3}\;.
\end{eqnarray}

\subsection{Tunneling couplings to the leads}
The electron tunnelling from  the left lead to the donor state and
from the acceptor state to the right lead are both given by the
Hamiltonian,
\begin{eqnarray}
H_{\rm tr} = - \sum_{k} T_{kL} \, a_D^{+} \,c_{kL} - \sum_{k}
T_{kR}\, c_{kR}^{+}\, a_A + {\rm H.c.}\;, \label{Htr}
\end{eqnarray}
where $ c_{k\alpha}^{+}, c_{k\alpha}$ are the electron creation and
annihilation operators, and $\alpha $ is the index for the leads.
The Hamiltonian of the leads is given by
 $$H_{LR} =
\sum_{\alpha} \varepsilon_{\alpha} n_{\alpha} \;\;\;\;\;\;{\rm
with}\;\;\;\;\;\;\; n_{\alpha} = \sum_{k}
c_{k\alpha}^{+}\,c_{k\alpha}\;.$$

\subsection{Thermal tunneling}
Activated by thermal fluctuations, electrons can tunnel between the
electron sites. Here, $H_{{\rm tun}}$, given by the following
expression,
\begin{eqnarray}
H_{\rm tun} = -\sum_{l}\Delta_{l,l'}\, a_{l}^{+}\,a_{l'} + {\rm
H.c.\;,}
\end{eqnarray}
which accounts for the thermal tunneling effects. Here
$\Delta_{l,l'}$ is the strength of the tunnelling coupling and the
\{$l,l'$\} indices refer to
 the pairs: \{D,~P\}, \{D,~P$^*$\}, \{A,~P\}, \{A,~P$^*$\}.

\subsection{ Light-induced excitations}
This part of the Hamiltonian accounts for the interaction of light
with the molecular triad and the antenna complexes. Under the
rotating-wave approximation, the light-induced excitation processes
can be described as
\begin{eqnarray} H_{\rm Light}~=~-~\sum_{k}F\,e^{i\omega_0
t}\,a_k^{+}\,a_{k^*} \ +\ {\rm H.c.}
\end{eqnarray}
where $k$ = P, An1, An2, An3, An4 and the field amplitude
$$ F={\cal E}_{\rm ext} \; d_{kk^*}\;,$$
where, $d_{kk^*}$ is the dipole moment.

\subsection{Coupling to a radiation heat bath and an Ohmic bath}
Coupling the system to a radiation heat bath causes radiation
leakage from the excited states. The following Hamiltonian accounts
for this radiation leakage
\begin{equation}
H_Q = - \sum_{\sigma}Q_{\sigma\sigma'} a_{\sigma}^{+}a_{\sigma'} +
{\rm H.c.}\;,
\end{equation}
where, $\{\sigma, \sigma'\}$ denotes the pairs of sites \{D,
P$^*$\}, \{A, P\}, \{P, P$^*$\}, \{An1,A n1$^*$\}, \{An2,
An2$^*$\}, \{An3, An3$^*$\}, \{An4, An4$^*$\}.

 The operators for the radiation bath,
\begin{eqnarray}
Q_{\sigma\sigma'} = e \ x_{\sigma\sigma'}\times {\cal E}_{\rm rad}
\end{eqnarray}
 are proportional to the projection of the fluctuating
electromagnetic field, ${\cal E}_{\rm rad}$, along the direction of
the corresponding dipole moment, $d_{\sigma\sigma'} = e\
x_{\sigma\sigma'}$.

The excited state of the photosensitive part of the molecular triad
 can be quenched by the electrode. Namely, lose the excitation
 energy when interacting with the electrodes.
We introduce $H_{\rm quench}$ to account for this energy-loss or
quenching processes.,
\begin{eqnarray}
 H_{\rm quench} = -Q_l \,a_{l}^{+}\,a_{l'} + { \rm H.c.}\;,
\end{eqnarray}
where, $Q_l$ is the variable of the Ohmic bath and $\{l, l'\} $ =
\{P, P$^*$\}.

\subsection{Interaction with the environment}
We have taken into account the effects of a dissipative environment
by the well-known system-reservoir model
\cite{bath1,FlagPRE08,JCP09Photo,JPCc09,PumpPRE08,MarcusSutin}.
\begin{eqnarray}
H_{\rm{env}}=&& \sum _{j}\left[ \frac{p_j^2}{2m_j}+
\frac{m_j\omega_j^2 }{2} \left( x_j +\frac{1}{2} \sum_i
x_{ji}\;n_i\right)^2 \right],\nonumber\\
\end{eqnarray}
where ${x_j,p_j}$ are the position and momentum  of the $j$th
oscillator with effective masses $m_j$ and frequencies $\omega_j$.
Here, $x_{ji}$ is a measure of the strength of the coupling between
the electron subsystem and the environment. We characterize the
phonon modes of the bath  by the spectral functions
$J_{ii'}(\omega)$, defined by
\begin{eqnarray}
J_{ii'} (\omega) = \sum_j \frac{m_j\omega_j^3
(x_{ji}-x_{ji'})^2}{2} \delta(\omega -  \omega _ j)\;.
\end{eqnarray}

The spectral function $J_{ii'}$ is related to the reorganization
energy $\Lambda_{ii'}$ for the $i \rightarrow i'$ transition,  by
the following equation:
\begin{eqnarray}
\Lambda_{ii'} = \int_{0}^{\infty} \frac{d \omega} {\omega}
J_{ii'}(\omega)
 = \sum_j \frac{m_j\omega_j^2 (x_{ji}-x_{ji'})^2}{2}.
\end{eqnarray}
\subsection{Unitary transformation} To remove the electron population
operators of the electron subsystem from $H_{\rm env}$, we use the
unitary transforation
 $\hat{U}=\prod _i \hat{U}_i$, where
\begin{eqnarray}
\hat{U}_i = \exp{\left[\frac{i}{2} \sum_{j} p_j x_{ji}n_i\right]},
\end{eqnarray}
The results of this unitary transformation are:
\newline
(i) When $\hat{U}$ is operated on an arbitrary function
$\Phi(x_j)$,
 a shift of the oscillator positions $x_j$ is produced.
\begin{eqnarray}
\hat{U}^{\dag}\Phi(x_j)\hat{U}=\Phi \left(x_j + \frac{1}{2}\sum_i
x_{ji}n_i\right).
\end{eqnarray}
\newline (ii) Another result of this unitary transformation is that all
 the transition operators acquire fluctuating factors.
\begin{eqnarray}
\hat{U}^{\dag} a_{i}^{\dag}a_{i'}\hat{U}=
e^{i\xi_{ii'}}a_{i}^{\dag}a_{i'}.
\end{eqnarray}

 The stochastic phase operators $\xi_{ii'}$ are given by
\begin{eqnarray}
 \xi_{ii'} = \xi_{i} - \xi_{i'}\;\;\;{\rm with}
\;\;\; \xi_i = \frac{1}{\hbar} \sum_{j} p_j x_{ji}\;.
\end{eqnarray}

\section{Master equations} The system under study can be
characterized by the  256 eigenstates of $H_0$. We expressed all
the operators described in previous section in the terms of the
density operators $\rho_{\mu \nu} \equiv |\mu \rangle \langle\nu|$.
To derive the time evolution of the diagonal elements $\rho_{\mu
\mu} \equiv \rho_{\mu}$ of the density matrix ($\rho_{\mu \nu}$),
we write the Heisenberg equation for the operators, with the
subsequent averaging $\langle \rho_{\mu}\rangle$ over the
environment fluctuations. Finally, we obtain the master equation
for the density matrix of the system
\cite{FlagPRE08,JCP09Photo,JPCc09,PumpPRE08,MarcusSutin},
\begin{eqnarray}\label{master}
\langle \dot{\rho}_{\mu}\rangle + \sum_{\nu}\gamma_{\nu \mu}
\langle \rho_{\mu}\rangle = \sum_{\nu}\gamma_{\mu\nu} \langle
\rho_{\nu}\rangle,
\end{eqnarray}
where, $\gamma_{\mu\nu}$ is the total relaxation matrix, which is
the sum of six types of relaxation rates:
\begin{eqnarray}\label{relax}
\gamma_{\mu \nu}= \gamma_{\mu \nu}^{\rm tr}+k_{\mu \nu}^{\rm
Forster}+k_{\mu \nu}^{\rm tun}+k_{\mu \nu}^{\rm light} + k_{\mu
\nu}^{\rm rad} + k_{\mu \nu}^{\rm quench}\;.
\end{eqnarray}
These relaxation rates will be described right below.

\subsection{ Electron tunneling rates between the leads and the molecular triad  }
 The first term of Eq.(\ref{relax}), $\gamma_{\mu \nu}^{\rm tr}$,
 represents the relaxation rates due the couplings of the triad to the L and
R electron reservoirs:
\begin{eqnarray}
\gamma_{\mu\nu}^{\rm tr} = \Gamma_L\left\{ |a_{D;\mu\nu}|^2 [ 1 -
f_L(\omega_{\nu\mu}) ]
 + |a_{D;\nu\mu}|^2 f_L(\omega_{\mu\nu})\right\} \nonumber\\
+ \Gamma_R\left\{ |a_{A;\mu\nu}|^2 [ 1 - f_R(\omega_{\nu\mu}) ] +
|a_{A;\nu\mu}|^2 f_R(\omega_{\mu\nu})\right\}\;,
\end{eqnarray}
where the $\Gamma_{\alpha}$ ($\alpha =$ L, R) are the resonant
tunneling rates. Here, the electron reservoirs have been
characterized by the Fermi distributions $f_{\alpha}(\omega) $,
\begin{eqnarray}
f_{\alpha}(E_{k\alpha})= \left[\exp\left(\frac{
E_{k_B\alpha}-\mu_{\alpha}}{T}\right)
 +1 \right]^{-1}.
\end{eqnarray}
with the temperature $T$ ($k_{\rm{B}} = 1, \hbar = 1$). The
electrochemical potentials $\mu_{L}$ and $\mu_R$ are the
controlling factors of the electron transition rates from the left
lead to the donor state, and from the acceptor state to the right
lead.

\subsection{ F\"{o}rster relaxation rates}
Here $k_{\mu \nu}^{\rm Forster}$ accounts for the excitation
transfer rates from the antenna complexes to the reaction center,
and also among the antenna complexes. The excitation transition
rates via the F\"{o}rster mechanism is given by
\begin{eqnarray}\label{Forster}
\kappa^{\rm Forster}_{\mu\nu} &=& \sum_{kl}
\sqrt{\frac{\pi}{\Lambda_{kl}T}} \;\; |V_{kl}|^2 \;\left[
|(a_{l}^{+}a_{l^*}a_{k^*}^{+}a_{k})_{\mu\nu}|^2\right. \nonumber\\
&+& \left.|(a_{l}^{+}a_{l^*}a_{k^*}^{+}a_{k})_{\nu\mu}|^2
\right]\exp \left[ - \frac{(\omega_{\mu\nu} + \Lambda_{kl} )^2}{4
\Lambda_{kl} T}
\right]\;,\nonumber\\
\end{eqnarray}
where $V_{kl}$ is the resonant F\"{o}rster relaxation rate and
$\Lambda_{kl}$ stands for reorganization energy. We denote, $V_{kl}
= V_F$ and $\Lambda_{kl}= \Lambda_{F}$ for any combinations of $k$
and $l$. The non-resonant
 exponential term of the above expression arises due to
the different energy gaps of the reaction center and the accessory
antenna complexes. Moreover,  the non-resonant exponential terms
depend on the reorganization energy.
\subsection{ Thermal tunneling rates}

The matrix element $k_{\mu \nu}^{\rm tun}$ of the Eq.~(\ref{relax})
are responsible for the relaxation processes arising from thermal
tunneling. These are given by
\begin{eqnarray}
\kappa^{\rm tun}_{\mu\nu} &=& \sum_{\sigma\sigma'}
\sqrt{\frac{\pi}{\Lambda_{\sigma\sigma'}T}} \;\;
|\Delta_{\sigma\sigma'}|^2 \;\left[
|(a_{\sigma}^{+}a_{\sigma'})_{\mu\nu}|^2\right. \nonumber\\ &+&
\left.|(a_{\sigma}^{+}a_{\sigma'})_{\nu\mu}|^2 \right]\exp \left[ -
\frac{(\omega_{\mu\nu} + \Lambda_{\sigma\sigma'} )^2}{4
\Lambda_{\sigma\sigma'} T} \right]\;,
\end{eqnarray}
where $\Delta$ is the resonant tunnelling rate,
  $\omega_{\mu\nu}$ is the energy difference between the states $\mu$ and $\nu$ (acting as a thermodynamic gradient), and
the reorganization energy $\Lambda$ are the main guiding factors of
the thermal tunneling rates $\kappa^{\rm tun}_{\mu\nu}$.

\subsection{Light-induced excitation rates}
 The  contribution $k_{\mu \nu}^{\rm light}$ to the total
relaxation matrix due to light-induced excitation processes is
\begin{eqnarray}
\kappa_{\mu\nu}^{\rm light} &=& \sum_{k}|F|^2 \sqrt{\frac{\pi}
{\Lambda_{\sigma \sigma^*}T}} \left\{|(a_{\sigma }^{+}a_{\sigma
^*})_{\mu\nu}|^2 \right. \nonumber
\\ &\times&\exp \left[ - \frac{(\omega_{\mu\nu}
+\omega_0 + \Lambda_{\sigma \sigma^*} )^2}{4 \Lambda_{\sigma \sigma^*} T} \right] \nonumber\\
&+& \left.|(a_{\sigma }^{+}a_{\sigma^*})_{\nu\mu}|^2  \exp \left[ -
\frac{(\omega_{\mu\nu} - \omega_0 + \Lambda_{\sigma \sigma^*}
)^2}{4
\Lambda_{\sigma \sigma ^*} T} \right]\right\} \nonumber\\
\end{eqnarray}
This rate includes contributions from the following transitions, P$
\to $P$^*$, An1 $\to$ An1$^*$, An2 $\to$ An$2^*$, An3 $\to$
An3$^*$, and An4 $\to$ An4$^*$.

\subsection{Relaxation rates due to radiation leakage}
Neglecting the effects of the environment on the radiation
transitions, $k_{\mu \nu}^{\rm rad}$ is given by
\begin{eqnarray}
\kappa_{\mu\nu}^{\rm rad} &=& \frac{2 n}{3}\; \sum_{\sigma
\sigma'}\; |d_{\sigma\sigma'}|^2\; [
|(a_{\sigma}^{+}a_{\sigma'})_{\mu\nu}|^2 +
|(a_{\sigma}^{+}a_{\sigma'})_{\nu\mu}|^2 ]\nonumber
\\
&\times& \left(\frac{\omega_{\mu\nu}}{c}\right)^3  \left[ \coth
\left( \frac{\omega_{\mu\nu}}{2T} \right) - 1 \right]\;,
\label{kapRad}
\end{eqnarray}
where, $n$ and $ d_{\sigma\sigma'}$ stand for the refraction index
 and the dipole moment, respectively.

\subsection{Lead-induced quenching rates of the excited states}
The last term of  Eq.~(\ref{relax}), $k_{\mu \nu}^{\rm quench}$,
describes the energy loss due to the quenching of the excited state
of the photosensitive part of the triad.
\begin{eqnarray}
\kappa_{\mu\nu}^{\rm quench} &=&\alpha_p
[|(a_{P}^{+}a_{P^*})_{\mu\nu}|^2 + |(a_{P}^{+}a_{P^*})_{\nu\mu}|^2
] \nonumber\\ &\times& \omega_{\mu\nu} \left[ \coth \left(
\frac{\omega_{\mu\nu}}{2T} \right) - 1 \right]\;.
\end{eqnarray}
%%%%%%%%%%%%%%%%%%%%%%%%%%%%%%%%%%%%%%%%%%%%%%%%%%%%%%%%%%%%%%%%%%%%
\section{Current and efficiency}

\subsection{ Electron current}

For weak couplings, the electron flowing between the leads and the
molecular triad is given by:
$$I_{e} = \left(\frac{d}{dt}\right)\sum_{k} \langle
c_{k\alpha}^{+}c_{k\alpha}\rangle\;,$$ We derive the equation of
the current in terms of the density matrix elements.
\begin{eqnarray}
I_{e} &=& \Gamma_R \sum_{\mu\nu} |a_{A;\mu\nu}|^2 [ 1 -
f_R(\omega_{\nu\mu}) ] \langle \rho_{\nu}\rangle \nonumber\\
&-& \Gamma_R \sum_{\mu\nu} |a_{A;\mu\nu}|^2 f_R(\omega_{\nu\mu})
\langle \rho_{\mu}\rangle \;.
\end{eqnarray}

\subsection{ Absorbed energy}
The total amount of energy absorbed per unit time ${\cal E}_{\rm
photon}$ by the molecular triad and antenna chromophores is
\begin{eqnarray}
{\cal E}_{\rm photon} &=& \sum_{\sigma}\omega_0 |F|^2\;
\sqrt{\frac{\pi}{\Lambda_{\sigma \sigma^*} T}} \; \sum_{\mu\nu}\;
|(a_{\sigma}^+a_{\sigma})_{\mu\nu}|^2 \;\langle \rho_{\mu} -
\rho_{\nu}\rangle \nonumber\\
&\times & \left( \exp\left[ - \frac{(\omega_{\mu\nu} -
\Lambda_{\sigma \sigma^*} +\omega_0)^2}{4\Lambda_{\sigma
\sigma^*}T}\right] \right.\nonumber
\\&-&
\left.\exp\left[ - \frac{(\omega_{\mu\nu} - \Lambda_{\sigma
\sigma^*} -\omega_0)^2}{4\Lambda_{\sigma \sigma^*}T}\right]
\right)\;,
\end{eqnarray}
where $\sigma$ = P, An1, An2, An3, An4.

\subsection{ Power-conversion efficiency}
 The power-conversion efficiency of the system is the
 the ratio of the output $E_{\rm output}$ and the input $E_{\rm input}$ energies,
\begin{eqnarray}\eta = \frac{E_{\rm output}}{E_{\rm input}} = \frac{{\cal E}_{\rm
pump}}{{\cal E}_{\rm photon}} = \frac{I_R (\mu_R -\mu_L)}{{\cal
E}_{\rm photon}}.
\end{eqnarray}
 The quantum yield is defined as
\begin{eqnarray}
\Phi = \frac{n_{\rm pump}}{N_{\rm photon}} = \frac{\eta
\;\omega_0}{V}\;.
\end{eqnarray}

\end{appendix}

%\textbf{Supporting Information Available online:} these
%supplementary materials describe the methods used in the paper.

\end{document}